\documentclass[fleqn,12pt]{article}
\usepackage{theorem}
\usepackage{amsmath}
\usepackage{amssymb}
\usepackage{amsopn}
\usepackage{latexsym}
\usepackage{setspace}
\usepackage{dsfont}
\usepackage{datetime}
\usepackage[allnumber,warning,easyold,math]{easyeqn}
\usepackage{color}
\usepackage[12pt]{moresize}
\usepackage{natbib}
\usepackage[bookmarks=true,bookmarksnumbered=true,pdfpagemode=None,pdfstartview=FitH,hidelinks]{hyperref}
\usepackage{fix-cm}
\usepackage[calcwidth]{titlesec}
\usepackage{fix-cm}
\usepackage{graphicx}
\usepackage{sgame}
\usepackage[font=footnotesize]{caption,subcaption}
\usepackage{multirow}

\theoremstyle{break}
\theorembodyfont{\itshape}

\usdate

\begin{document}

\linespread{1}

\title{\textbf{\Large Status Quo Bias and the Decoy Effect:\\ A Comparative Analysis in Choice under Risk}\thanks{The authors (ordered alphabetically) thank Erik Eyster, Pietro Ortoleva and Charlie Sprenger for helpful comments and Laure-Aliz\'{e}e Le Lannou, Zhibo Xu for research assistance. Gerasimou gratefully acknowledges grants from the British Academy and the Carnegie Trust.}}

\author{Miguel Costa-Gomes\thanks{School of Economics \& Finance, University of St Andrews.} \and Georgios Gerasimou\thanks{School of Economics \& Finance, University of St Andrews. Corresponding author: \href{mailto:gg26@st-andrews.ac.uk}{\color{blue}gg26@st-andrews.ac.uk\color{black}}.}}

\linespread{1.1}
\parskip=5pt

\maketitle

\begin{abstract} 
Inertia and context-dependent choice effects are well-studied classes of behavioural phenomena. While much is known about these effects in isolation, little is known about whether one of them ``dominates'' the other when both can potentially be present. Knowledge of any such dominance is relevant for effective choice architecture and descriptive modelling. We initiate this empirical investigation with a between-subjects lab experiment in which each subject made a single decision over two or three money lotteries. Our experiment was designed to test for dominance between \textit{status quo bias} and the \textit{decoy effect}. We find strong evidence for status quo bias and no evidence for the decoy effect. We also find that status quo bias can be powerful enough so that, at the aggregate level, a fraction of subjects switch from being risk-averse to being risk-seeking. Survey evidence suggests that this is due to subjects focusing on the maximum possible amount when the risky lottery is the default and on the highest probability of winning the biggest possible reward when there is no default. The observed reversal in risk attitudes is explainable by a large class of K\"{o}szegi-Rabin (2006) reference-dependent preferences.
\end{abstract}

\thispagestyle{empty}

\vfill
\pagebreak

\parindent=15pt

\linespread{1.4}

\setcounter{page}{1}

\section{Introduction}

\textit{Status quo bias} \citep{samuelson&zeckhauser} is the phenomenon whereby decision makers are significantly more likely to choose a market alternative such as a retirement savings plan or energy tariff when it is the ``default'' option than when it is not. The \textit{decoy effect} \citep{huber&payne&puto} refers to the tendency of decision makers to exhibit a specific change in their choice between two alternatives when also presented with a third one that is dominated by only one of the original two. These phenomena have been documented empirically in a wide range of choice environments and are among the most well-known behavioural ``anomalies'' that cannot be explained by standard models of rational choice. Accordingly, they have also been the subject of extended interdisciplinary modelling, with a plethora of suggested explanations.

Although much is known about these effects individually little is known about which one of them --if any-- is \textit{stronger}. Knowing the answer to this question is important for at least two reasons:
\begin{enumerate}
	\item \textit{Choice architecture} \citep{thaler-sunstein-balz13,johnson-goldstein03,madrian&shea}. Policy makers or physical/online retailers may opt for menu designs that aim to influence choice --e.g. over (risky) retirement savings plans or (riskless) student meals, consumer products, or even flu vaccination schedules \citep{maltz-sarid20}-- in a certain direction. Although the significance of both status quo bias and, separately, the decoy effect on such choice architecture is well-known, how decision makers react in a setting where both effects could potentially be present, is an open question, that is yet to be addressed. This question has very important practical implications in choice architecture, such as the feasibility of a menu designer to ``nudge'' a decision maker away from some existing default option by introducing a decoy dominated option for the competitor target alternative, for example.
	\item \textit{Modelling guidance.} While numerous models that explain one or more of these effects are now available, no experimental or empirical study that we are aware of has guided bounded-rational choice modelling of this kind at such a more refined level. Knowing whether inertia or some context-dependent choice effect is more powerful than the other --and if so, in which domains this is true-- is clearly essential to provide precise empirical foundations for such modelling. Indeed, one of the criticisms of some models of bounded rationality is that often capture just one bias. \color{black} But, for them to capture several biases we will need to know what, if any, additional features of the decision-making situation matter in assessing the relative strength of such biases. \color{black}
\end{enumerate}  

Motivated by the above, in this paper we initiate the comparative empirical analysis of inertia and context-dependent choice using data from a novel between-subjects experiment that features \color{black} a single decision by each subject in each of its three treatments.  The subjects' decision was made from a set of two or three lotteries each, with three strictly positive monetary outcomes. The aim is to understand whether in a decision environment of choice under risk where status quo bias and the decoy effect could potentially influence choices the former is stronger or weaker than the latter. \color{black} We find that: 
\begin{enumerate}
	\item Status quo bias unambiguously prevails over the decoy effect and, contrary to the findings reported in other studies on choice under risk, the latter effect is completely absent in our data.
	\item The riskier of the two main lotteries is chosen significantly more often when it is the status quo than when it is not, due to subjects giving more significance to the dominant ``maximum reward'' dimension of that lottery in this case.
	\item This status-quo induced reversal in the risk attitude of the average subject is explainable by a large class of K\"{o}zsegi-Rabin (\citeyear{koszegi&rabin}) reference-dependent preferences.
\end{enumerate}

\color{black} We wish to stress from the outset that a main feature of our design is that subjects were asked to make a single choice. Out of a single menu of lotteries they choose one that was then played out to determine their total rewards. \color{black} This simplicity made the experimental task and its incentivization very easy \color{black} for subjects \color{black} to understand, \color{black} while it presumably also maximized their focus, motivation and quality of that unique choice. \color{black} Moreover, the fact that subjects saw and made a choice from a single menu enables us to rule out any linkages between decisions across different problems that sometimes arise in within-subjects designs.

\section{Experiment}

The experiment was conducted at the University of St Andrews Experimental Economics laboratory between May 2018 and April 2019. Participants were recruited with ORSEE \citep{ORSEE15} and the experimental interface was created in z-tree \citep{ztree}. Because we used a printed end-of-experiment survey form and a non-computerized way to randomly determine the subjects' winnings from their chosen lottery, the average total duration of each session was 35 minutes. In addition to their lottery winnings, subjects received a \pounds 2 participation fee.

\begin{figure}[htb]
	\caption{The three lotteries}
	\label{fig:lotteries}
	\centering 
	\begin{subfigure}{0.32\textwidth}
		\includegraphics[width=\linewidth]{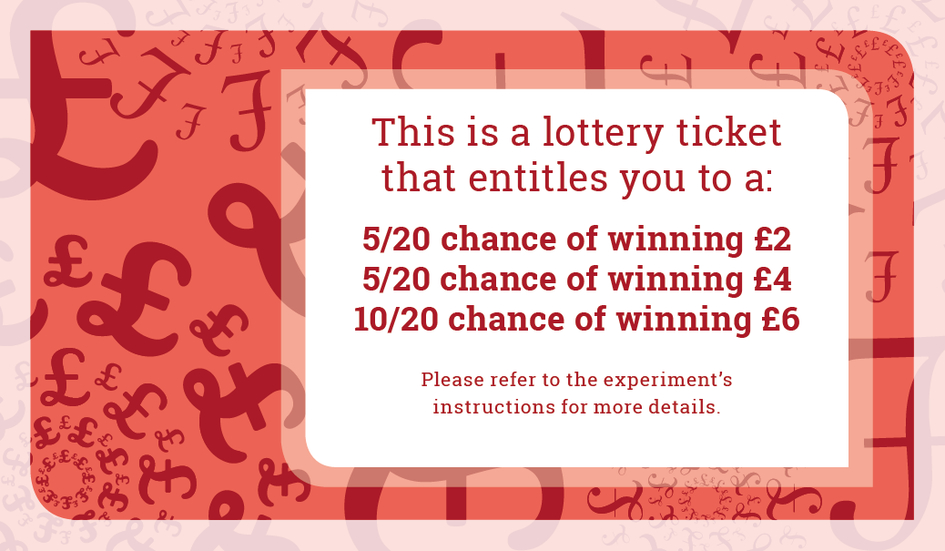}
		\caption{Lottery \textit{A}}
		\label{fig:A}
	\end{subfigure}\hfil
	\begin{subfigure}{0.32\textwidth}
		\includegraphics[width=\linewidth]{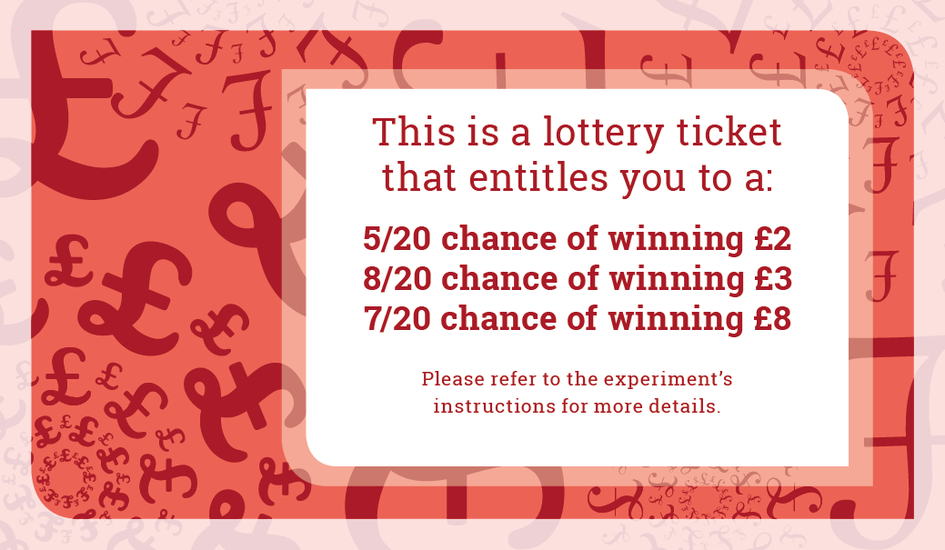}
		\caption{Lottery \textit{B}}
		\label{fig:B}
	\end{subfigure}\hfil 
	\begin{subfigure}{0.32\textwidth}
		\includegraphics[width=\linewidth]{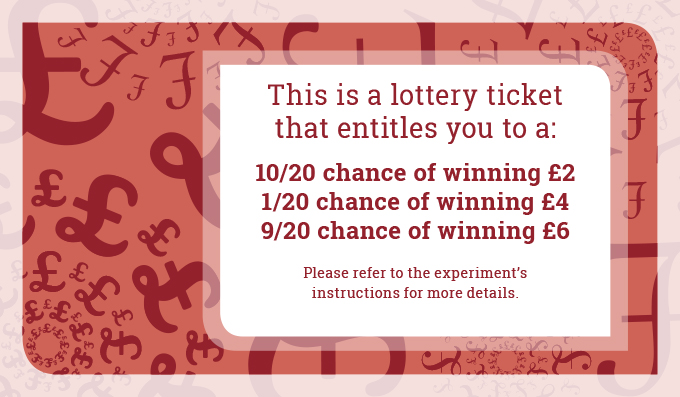}
		\caption{Lottery \textit{A-decoy}}
		\label{fig:Adec}
	\end{subfigure}
\end{figure}

\begin{figure}[!htbp]
	\caption{The three treatments}
	\label{fig:treatments}
	\centering 
	\begin{subfigure}{1\textwidth}
	\caption{Treatment 1: No decoy or default lottery}
		\includegraphics[scale=0.5]{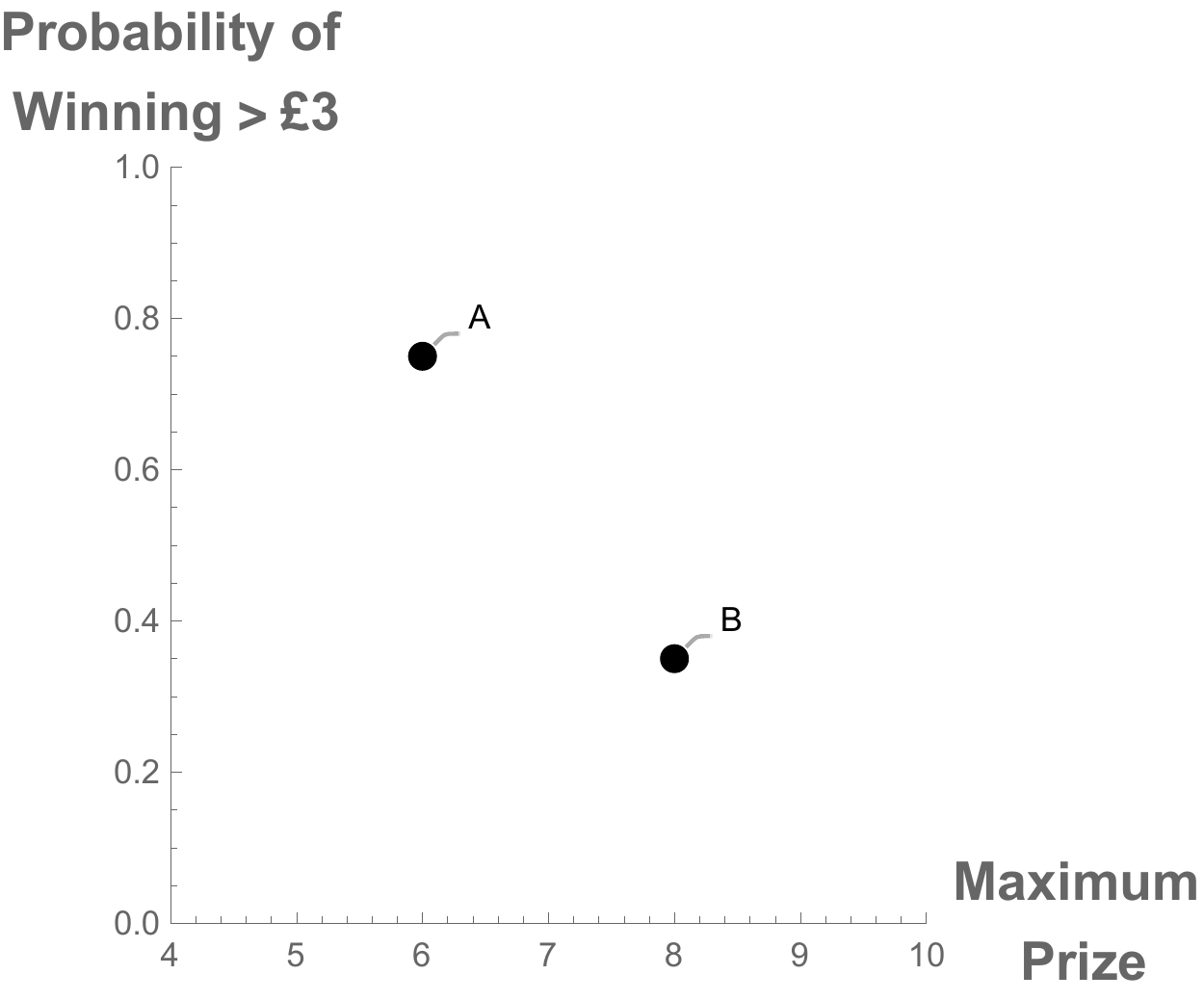} \hspace{50pt}
		\includegraphics[scale=0.30]{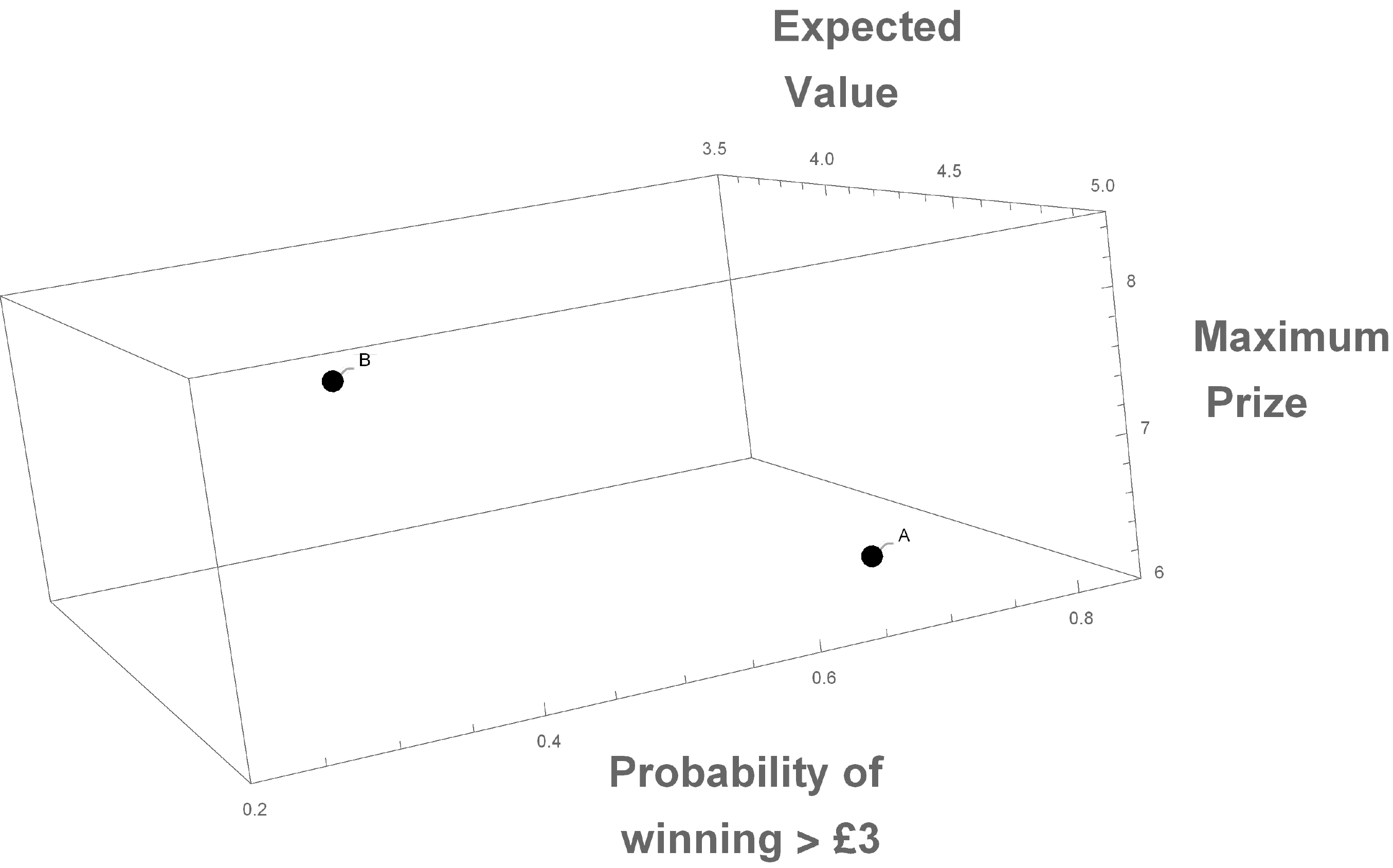}
		\label{fig:treat1}
	\end{subfigure}\hfil\vspace{30pt}

	\begin{subfigure}{1\textwidth}
		\caption{Treatment 2: Decoy and no default lottery}
		\includegraphics[scale=0.5]{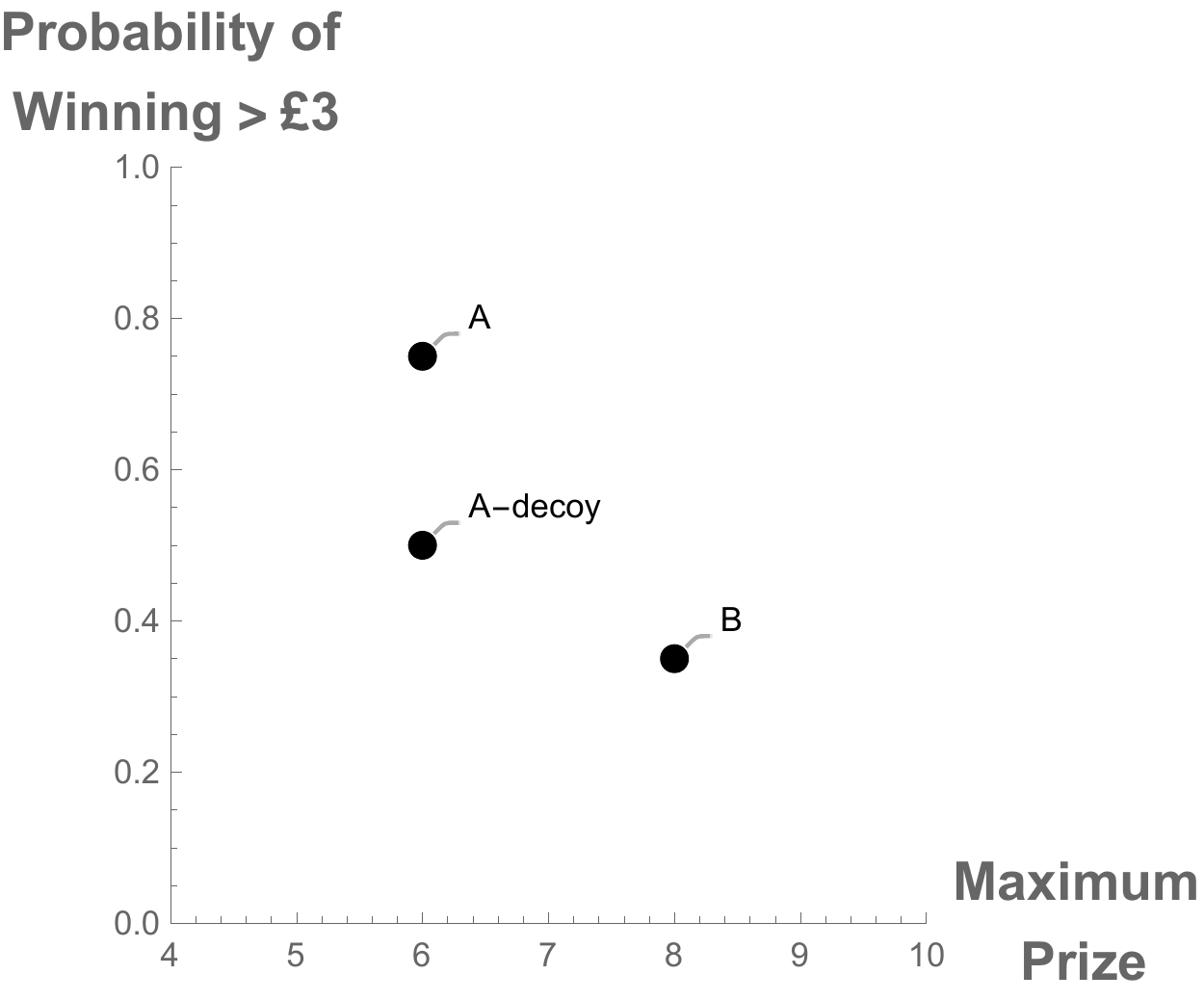} \hspace{50pt}
		\includegraphics[scale=0.30]{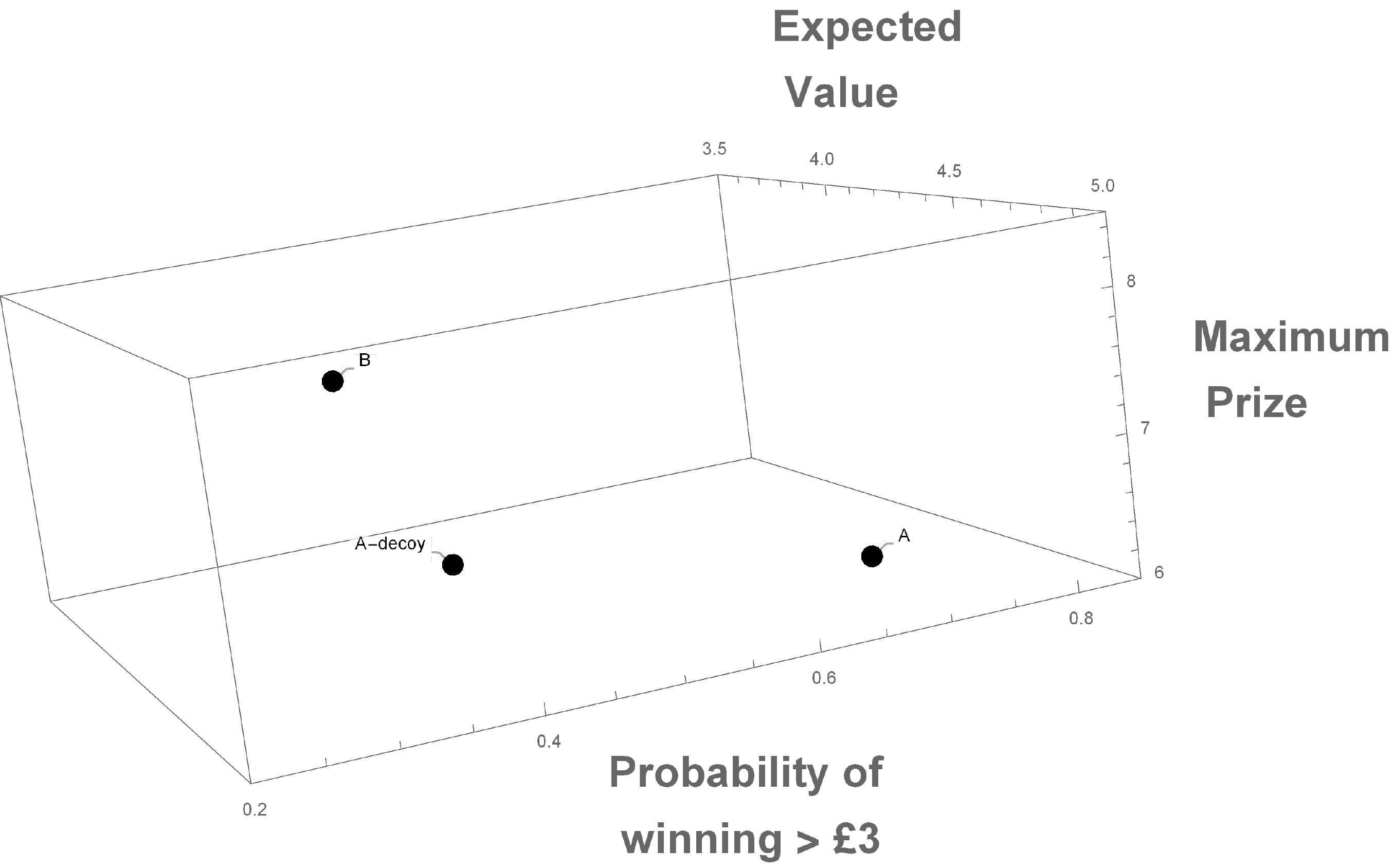}
		\label{fig:treat2}
	\end{subfigure}\vspace{30pt}

	\begin{subfigure}{1\textwidth}
		\caption{Treatment 3: Decoy and default lottery (\textit{B})}
		\includegraphics[scale=0.5]{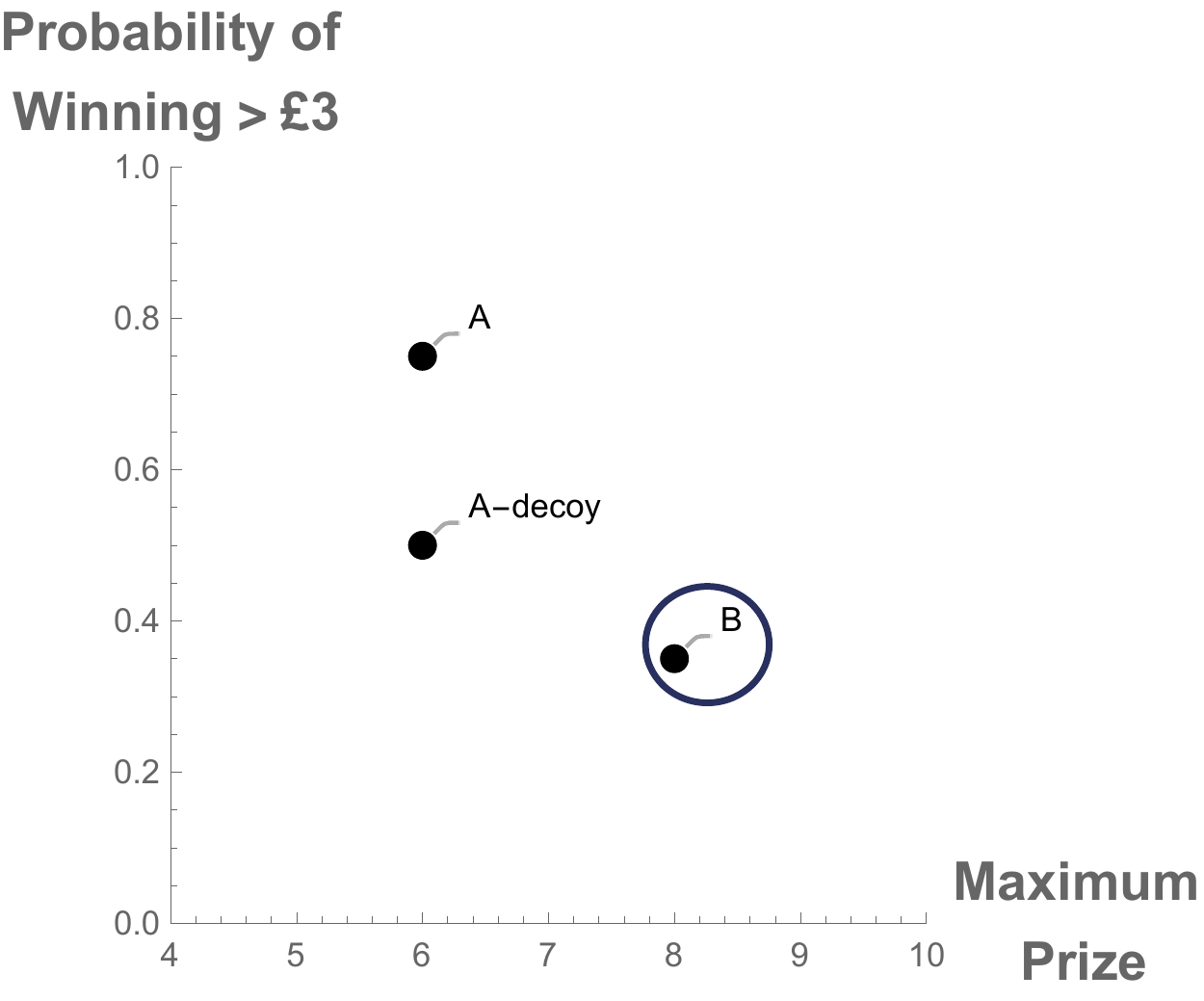}\hspace{50pt}
		\includegraphics[scale=0.30]{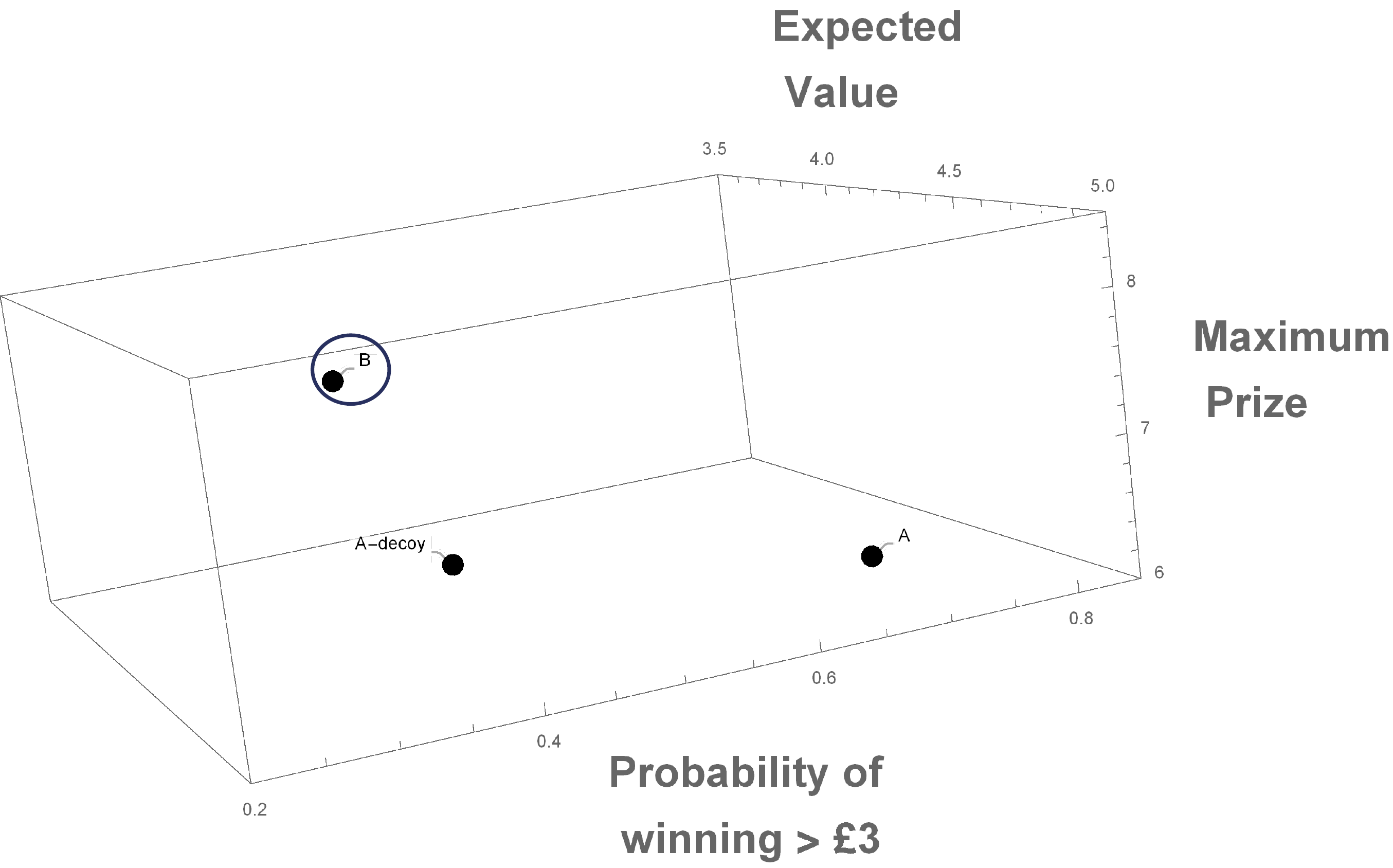}
		\label{fig:treat3}
	\end{subfigure}
\end{figure}

Figures \ref{fig:lotteries} and \ref{fig:treatments} present, respectively, the three lotteries and treatments/decision problems that we employed. In the first treatment, subjects were asked to choose between lotteries \textit{A} and \textit{B} without being initially endowed with either of them. In the second treatment, they were asked to choose between lotteries \textit{A}, \textit{A-decoy} and \textit{B}, also without an initial endowment. In the third treatment, they were given an envelope that contained a printed card of lottery \textit{B}. Once everyone received their envelope, they were instructed to open it and read that card. Two minutes later, the experimental interface was activated and presented the menu comprising lotteries \textit{A}, \textit{A-decoy} and \textit{B} on their screen, inviting subjects to make a choice from it. The menu order and wording in this decision problem were identical to those in the second treatment and, as far as lotteries \textit{A} and \textit{B} are concerned, the first treatment too. Subjects in this treatment were told from the beginning that if they chose a different lottery to the one they had been endowed with, the experimenters would come to their desks and change their printed lottery card with the one corresponding to their chosen lottery. Our method, therefore, imposes an exogenous status quo and differs from within-subject designs where this emerges endogenously as the subjects' choice in an initial decision problem.

\begin{table}[!htbp]
	\centering \footnotesize
	\caption{Attribute dominance within the three pairs of lotteries.}
	\renewcommand{\arraystretch}{1.1}
	\makebox[\textwidth][c]{
	\begin{tabular}{|l|c|c|c|c|c|c|}
		\hline
								& \textbf{First-order}		& \textbf{Second-order}   & \textbf{Expected}  & \textbf{Maximum} & \textbf{Probability of}   & \textbf{Probability of} \\  
								& \textbf{stochastically}  	& \textbf{stochastically} & \textbf{value}	   & \textbf{prize}   &  \textbf{winning}   & \textbf{winning the} \\ 
								& \textbf{dominant}			& \textbf{dominant}       &					   & 				  & \textbf{more than \pounds 3}  & \textbf{\color{black} lottery's}\\
\textit{Lottery pair}	& 							& 					      &					   & 				  &   &  \textbf{largest prize} \\
\hline
\textbf{\textit{A vs B}}        & None 						& \textit{A}			  & Equal		 	   & \textit{B}  	  & \textit{A}	& \textit{A} \\ 
\hline
\textbf{\textit{A vs A-decoy}}  & \textit{A}  				& \textit{A}			  & \textit{A}		   & Equal 			  & \textit{A}	& \textit{A} \\ 
\hline
\textbf{\textit{B vs A-decoy}}  & None						& None 					  & \textit{B} 	       & \textit{B}  	  & \textit{A-decoy}	& \textit{A-decoy} \\
\hline
	\end{tabular}
}
	\label{tab:dominance}
\end{table}

Table \ref{tab:dominance} summarizes the dominance relationships within each of the three pairs of lotteries that are derived from \textit{A}, \textit{B} and \textit{A-decoy}. In particular: 
\begin{enumerate}
	\item Lottery \textit{A} first- and second-order stochastically dominates \textit{A-decoy}.
	\item Lottery \textit{B} offers a higher expected value and a higher maximum possible reward than \textit{A-decoy}, but a lower probability of winning that maximum reward. In fact, \textit{B} offers a lower probability than both \textit{A} and \textit{A-decoy} of winning more than \pounds 3.
	\item Lottery \textit{B} is a mean-preserving spread of (hence more risky than) \textit{A} but offers a higher maximum possible reward.
\end{enumerate}
This summary clarifies that \textit{A} does indeed asymmetrically dominate \textit{A-decoy} in menu $\{$\textit{A}, \textit{B}, \textit{A-decoy}$\}$, and that \textit{A} and \textit{B} cannot be ranked in an unambiguous manner even though, of course, every \textit{risk-averse} expected-utility maximizer would choose the former \citep{hadar-russel69}.
The three-dimensional graphs in Figure \ref{fig:treatments} (right) illustrate the dominance relationships reported in Table \ref{tab:dominance} based on the information in the penultimate three columns: ``Expected value''; ``Maximum prize''; ``Probability of winning more than \pounds 3''. The two-dimensional graphs (left) simplify this presentation by focusing on the latter two dimensions. All graphs are to scale.

We stress that, unlike previous experimental studies \citep{herne98,masatlioglu&uler} where the asymmetrically dominated decoy and status-quo options coincided, our study appears to be the first where individuals were faced with the problem of deciding between keeping their default option and changing it for a different one that itself asymmetrically dominates another option. In conjunction with the other two standard treatments, this novelty of our design is precisely what enables one to compare the relative strengths of status quo bias and the decoy effect and understand which one prevails over the other, if any. 

We also remark that a fourth treatment should also be included in general when such a relative-strength question is raised concerning two phenomena. This treatment would feature the decision problem where subjects are presented with the binary menu after being initially endowed with lottery \textit{A}. This, in particular, would allow for a double comparison of the two effects' relative magnitude. The first would be the direct one that is based on our last treatment. The second would be an indirect comparison where the choice probabilities of lottery \textit{A} would be compared relative to the neutral binary treatment, both when \textit{A} is the asymmetrically dominant option and when it is the default option in that new binary environment. In light of the results presented below, we judged that this treatment was redundant for the purposes of this study, and we therefore did not introduce it. For the same reason we also did not conduct sessions for the symmetric versions of the second and third treatments where the asymmetrically dominant and default options are \textit{B} and \textit{A}, respectively.

As far as the implementation is concerned, we note that once the subjects in all treatments confirmed their choices, they were individually invited to draw a ball from an urn with twenty numbered balls. The number on the randomly selected ball determined the prize won by that subject in a way that was true to the chosen lottery's description. In the case of lottery \textit{A}, for example, the prize was \pounds 2 if the number was 1--5, \pounds 4 if it was 6--10, and \pounds 6 if it was 11--20. This had been pre-specified and communicated to subjects before the process began. Finally, subjects were given a printed questionnaire where they were asked to enter their subject ID and provide information about their field and level of study and explain how they made the decision on which lottery to choose. Our sample comprised primarily undergraduate and postgraduate students of many fields. There were no significant differences in student characteristics across the three treatments, either between graduate/undergraduate or between economics/non-economics students. Economics/finance/management students, in particular, comprised less than a third of participants in each treatment.

\section{Results}

Our choice-based findings are summarized in Table \ref{tab:results1}. In the neutral binary-menu treatment, two thirds of all subjects chose the safer lottery \textit{A}, and the difference in the two proportions was highly significant. Remarkably, and against the decoy effect, introducing the asymmetrically dominated lottery \textit{A-decoy} in the second treatment actually led to a \textit{decrease} in the choice probability of the target lottery \textit{A}. Although the latter was favoured by just over half of the subjects in this treatment, the difference in the two proportions is insignificant, as is the difference in the proportions of subjects choosing it in Treatments 1 and 2. By contrast, we find a very significant status quo bias in the third treatment where \textit{B} is the default and \textit{A} the asymmetrically dominant option, with a dramatic reversal in choice probabilities where two thirds of all subjects chose to keep \textit{B} despite the fact that it is a mean-preserving spread of \textit{A}. The choice probabilities of \textit{B} are significantly different across Treatments 1 and 3, while those of \textit{A} are also significantly different across Treatments 2 and 3. Together, these observations point to the conclusion that status quo bias exists and prevails over the (absent) decoy effect in our data. Finally, under the maintained assumption that the subject pools are random samples that come from the same population, our findings suggest that status quo bias could potentially be altering the risk attitude of a proportion as high as $\frac{0.66-0.37}{0.66}\approx 0.44$ of the risk-averse decision makers in the neutral binary treatment and turn them into risk-seeking ones when \textit{B} is the default option.

\begin{table}[!htbp]
\footnotesize
\captionsetup{justification=centering}
	\caption{\color{black} Top panel: choice probabilities in the three treatments.\\ Bottom panel: treatment-effect tests.}
\renewcommand{\arraystretch}{1.1}
\makebox[\textwidth][c]{
\begin{tabular}{|r|c|c|c|}
		\hline
	\multirow{2}{*}				& {\textbf{Treatment 1}} 	 & \textbf{Treatment 2}			 				 		 		  & \textbf{Treatment 3}\\ 
 								& 	(binary)	 			 &  (decoy) 													  & (decoy \& default)\\ 
\hline
\textit{A}          			&  0.66				     	 & 0.54 						 				 		 		  & 0.37  \\ 
\hline
\textit{B}          			&  0.34		             	 & 0.45 						 				 		 		  & 0.61 \\
\hline
\textit{A-decoy}          		&  --		             	 & 0.01 						 				 		 		  & 0.02 \\
\hline
$n=$ 		          			&	74						 &	78							 						 		  & 59 \\
\hline
 $p$-value  					& 0.007	   					 &   0.571											  			  & 0.067 \\
\hline
\hline
								& \textbf{Treatments 1 \& 2} & \textbf{Treatments 1 \& 3} 	 								  & \textbf{Treatments 2 \& 3}\\ 
								& (on \textit{A})		 	 & (on \textit{B})		 									  	  & (on \textit{A})\\
\hline
$p$-value 	                	&	0.138 		             & 0.003						 								  &   0.059\\
\hline
\end{tabular}
}
{\footnotesize \color{black} Notes: (i) top-panel $p$-values from 2-tailed binomial tests with an equal-proportions null; (ii) bottom-panel $p$-values from 2-tailed Fisher's exact tests.}
\label{tab:results1}
\end{table}

We now move on to the analysis of end-of-experiment survey responses and, in particular, on the subjects' own explanation on the reason why they chose that particular lottery in their respective treatment. Responses that contained logical errors (e.g. \textit{``because it had a higher expected value''} when in fact it did not) or those that were otherwise unclear were dropped from this analysis.
Table \ref{tab:responses} presents the main groups into which the subjects' responses were categorized, and how the strength of these groups varied with the treatment. In the two primary groups, the subjects reported as their main criterion the ``highest probability of the largest reward'' (in which \textit{A} dominates \textit{B}) and ``the largest possible reward'' (in which \textit{B} dominates \textit{A}). A few subjects --primarily economics/finance students-- either supplemented these responses with references to being ``risk-averse'' or ``risk-seeking'', respectively, or simply mentioned those risk attitudes directly and without any further explanation. Although the corresponding matching responses were grouped together in the table, we stress that most subjects came from disciplines that would not have familiarized them with the decision-theoretic terminology for attitudes to risk, and yet they could consistently be grouped into one of these categories. Notably, a few subjects (mainly in the binary treatment) cited ``the largest probability of the middle reward'' and ``the largest middle reward'' as their primary criterion. 

\begin{table}[!htbp]
	\centering\footnotesize
	\captionsetup{justification=centering}
	\caption{Categorization of the subjects' choice justification based on survey data.}
	\begin{tabular}{|l|c|c|c|}
		\hline
		& \textbf{Treatment 1}  			& \textbf{Treatment 2}  			& \textbf{Treatment 3} \\
		\textbf{Reason for choice}  & (binary)							& (decoy)							& (decoy \& default)\\
		
		\hline
		``Risk averse'' or ``opted	   	&&& \\
		for the highest probability &55.2\% 								&45\% 								&32\%\\
		of the largest reward''		&&&\\
		\hline
		``Risk seeking'' or ``opted		& 									&	 								&\\
		for the largest 			&31.3\%								&53\%								&68\%\\
		possible reward''				&&&\\
		\hline
		``Opted for the				&4.5\% 								&2\% 								&--\\
		largest middle reward''		&&&\\
		\hline
		``Opted for the				&&&\\
		highest probability 		& 3\%								&--									&--\\
		of the middle reward''		&&&\\
		\hline
		 Other reason 				& 								&--									&--\\
		(e.g. ``expected value of	&  6\% &&\\
		maximum outcome'')			&&&\\
		\hline
		Evaluable responses			& 67/74								&60/78								&47/59\\ 
		\hline
	\end{tabular}
	\label{tab:responses}	
\end{table}

All subjects were allowed to use the calculator that was embedded in the experimental interface or to use pen and paper for calculations. Most subjects did so, and many mentioned as part of their response that they had worked out the equality in the expected values of \textit{A} and \textit{B}. This suggests that the expected-value calculation might have been the primary criterion of choice for those subjects.

As far as treatment variation is concerned, the main interest lies again in the comparison between the neutral binary treatment and the decoy \& default treatment. Consistent with the observed choice-based reversal in risk attitudes, this comparison suggests that the act of endowing subjects with the default lottery \textit{B} had a clear effect in making the ``largest reward'' criterion important to 68\% of evaluable subjects in the last treatment from 31\% in the neutral binary treatment ($p$-value from two-tailed Fisher exact test: $<0.001$). Notably, only one subject in the last treatment mentioned the ``convenience'' of choosing their default lottery. These facts suggest that the majority of subjects who kept their default lottery did so following an active deliberation on whether to change it for the alternative undominated lottery, and the outcome of this deliberation was determined in favour of the relative advantage of the default lottery.

\section{Discussion}

The observed reversal in risk attitudes at the aggregate data level  is incompatible with the expected-utility model. Indeed, any risk-averse (risk-seeking) decision maker whose behaviour is captured by that model would choose lottery \textit{A} (\textit{B}) in all treatments. Yet, no such decision maker would choose \textit{A} in the first two and \textit{B} in the third treatment, which our results suggest that a sizeable fraction of individuals would do. In principle, one could assess this further using a within-subjects experimental design. While data elicited in this way could indeed provide additional information at the level of the individual decision maker, an important and well-known challenge associated with such elicitations is that subjects' responses could potentially be \color{black} sensitive to the order in which problems are presented or \color{black} biased due to reduction of cognitive dissonance, which in this context would manifest itself in the need for later choices to be consistent with earlier ones.

Following \cite{kahneman-tversky79}, \cite{tversky-kahneman92} and the ensuing literature, a natural approach to explain this aggregate-level behaviour is to employ a model of reference-dependent preferences. In particular, given the probabilistic nature of the reference point/status quo in the third treatment, a potentially suitable model in this class is the one due to \cite{koszegi&rabin} which, unlike original and cumulative prospect theory, is well-defined in cases where the reference point is possibly stochastic.

Under the K\"{o}szegi-Rabin model, the expected payoff from lottery $P$ when the reference point is another lottery $Q$, and when both $P$ and $Q$ have a finite support with the same cardinality $k$, is given by 
\begin{equation}
	\label{KR} 
	U_{KR}(P|Q):=\sum_{j=1}^k Q_j\left[\sum_{i=1}^k P_i\cdot u(x_i|r_j)\right],
\end{equation}
where  
\begin{equation}
	\label{reference}
	u(x|r):=m(x)+\mu\big(m(x)-m(r)\bigr).
\end{equation}
In this model, $m(\cdot)$ captures consumption utility whereas $\mu(\cdot)$ captures gain-loss utility relative to the reference value $r$. As in much of the analysis in \cite{koszegi&rabin} and other studies (e.g. Sprenger, \citeyear{sprenger15}; Masatlioglu and Raymond, \citeyear{masatlioglu-raymond16}), we assume that $\mu(\cdot)$ is piecewise-linear, with
\begin{equation}
\label{piecewise}
\mu(z):=
\left\{
\begin{array}{ll}
z, & \text{if $z\geq 0$}\\
\lambda \cdot z, & \text{if $z<0$},
\end{array}
\right.
\end{equation}
where $\lambda>0$ is the coefficient of loss aversion.

Recalling now that the show-up fee in our experiment was \pounds 2, it may be plausible to think of $r= 2$ in \eqref{reference} as the deterministic reference point in the first and second experimental treatments, and to augment the lottery prizes by \pounds 2 in order to translate them into the possible final wealth levels at the end of the experiment. It may also be plausible to think of lottery \textit{B} as the stochastic reference point in the third treatment, and to also augment the three prizes featured in it by the riskless \pounds 2 show-up fee. Under this formulation, the K\"{o}szegi-Rabin model can explain our findings if there exist $m(\cdot)$ and $\lambda>1$ such that
\begin{eqnarray}
\label{eq1} U_{KR}(A\oplus 2\,|\,2)   & > & U_{KR}(B\oplus 2\,|\, 2)\\
\label{eq2} U_{KR}(B\oplus 2\,|\,B\oplus 2) & > & U_{KR}(A\oplus 2\,|\,B\oplus 2) 
\end{eqnarray}
where $P\oplus 2$ means that each possible prize in lottery \textit{P} has been scaled up by \pounds 2. Under the specification \eqref{KR}--\eqref{piecewise}, inequalities \eqref{eq1} and \eqref{eq2} are indeed simultaneously satisfied for \textit{any} loss-aversion parameter $\lambda>1$ and \textit{any} CRRA consumption utility function defined by $m(z):=\frac{z^{1-\rho}}{1-\rho}$ or $m(z):=\log(z)$ for $\rho\neq 1$ and $\rho=0$, respectively, as well as \textit{any} CARA consumption utility function defined by $m(z):=1-e^{-Az}$ for $A>0$. Therefore, this status-quo induced reversal in risk-attitudes can be accommodated by a very large class of K\"{o}szegi-Rabin preferences.

Other experimental studies on choice over binary money lotteries or uncertain acts that also investigated the effect of default alternatives on choice include \cite{samuelson&zeckhauser}, \cite{bar-hillel&neter}, \cite{roca&hogarth&maule}, \cite{sprenger15}, \cite{dean-kibris-masatlioglu17}, \cite{maltz&romagnoli} and Chapman et al. (\citeyear{chapman_etal19}). Consistent with our findings, these papers also documented a status quo bias. In their within-subject experiments, in particular, \cite{sprenger15} and Chapman et al. (\citeyear{chapman_etal19}) reported a variety of risk-attitude reversals which are explainable by the \cite{koszegi&rabin} model and hybrid models of reference-dependent preferences, respectively. A contribution of our analysis to this literature is that we find clear evidence that status quo bias can overturn the decision makers' attitude toward risk and make them risk-seeking even when the relevant lotteries feature three outcomes that generate second-order stochastic dominance vs. maximum-reward trade-offs which cannot arise with simple binary lotteries.

Contrary to our results, however, studies that include \cite{herne99}, \cite{soltanietal12} and \cite{castillo20} found evidence for a decoy effect on choice under risk. Unlike our experiment that was based on lotteries with three non-zero monetary outcomes and allowed for studying the role that the above novel trade-offs may have on the incidence of this effect, the lotteries in all these studies featured simple binary gambles with a single non-zero outcome. \cite{herne99} and \cite{soltanietal12} used within-subject experimental designs and reported decoy effects for a statistically significant minority of subjects. In the latter study, the authors also found evidence for the decoy effect even when the decoy lottery was shown to subjects but was not available to choose. \cite{castillo20} reported on between- as well as within-subject experiments with such lotteries and found evidence to suggest that the decoy effect was present in his data, but in half the magnitude that had been reported in previous studies. 

Taking these facts into account, \color{black} and considering also the data from a similarly structured pilot experiment\footnote{This experiment was conducted in April-May 2017, also at the University of St Andrews Experimental Economics Lab. See the Appendix for more details.} with different lotteries that point in the same direction, \color{black} we conjecture that the complete absence of the decoy effect in our study is potentially driven by the fact that the three lotteries featured three outcomes, and this made it harder for subjects to detect the asymmetric dominance relationship in the relevant menu. At the same time, the fact that only one subject ($\approx$ 1.5\%) in each of Treatments 2 and 3 chose the dominated lottery means that a \textit{similarity effect} \citep{tversky72} is not a likely explanation for the absence of the decoy effect, because this would predict comparable shares for both \textit{A} and \textit{A-decoy}. \color{black} It is possible, however, that the dominance of $B$ over both $A$ and $A$-\textit{decoy} in the ``maximum prize'' dimension neutralized the dominance of $A$ over $B$ and $A$-\textit{decoy} in the probability of winning more than $\pounds 3$ and the probability of winning the lottery's largest prize. Such a potential neutralizing effect may have steered more subjects towards $B$ at $\{A,$$A$-\textit{decoy}$,B\}$ than at $\{A,B\}$. This effectively suggests that $A$-\textit{decoy} may have acted as a decoy for $B$ rather than $A$ for those subjects. While we believe that this possible explanation is worthy of further study and verification in future research, we note that in any case our finding points toward some potential limits of the decoy effect in choice under risk that economists and behavioural scientists should be aware of. Similar limits have also been documented recently in the consumer psychology and marketing literatures in \cite{frederick-lee-baskin14}, \cite{yang-lynn14} and \cite{trendl-stewart-mullett21}, who found no attraction effects in environments of naturalistic choice where the alternatives were presented without any numerical information attached to their description.

\section{Concluding Remarks}

While inertia and context-dependent choice effects are well-documented and influential behavioural phenomena, no study that we are aware of has attempted to compare their relative strength. Knowing which phenomenon, if any, is dominant in this sense is important for choice-architecture and descriptive-modelling purposes. In this paper we initiate the study of such comparisons by means of a novel experimental design that allowed for a direct such test between status quo bias and the attraction/asymmetric dominance/decoy effect in an environment of choice under risk. Our primary findings are that status quo bias prevails over the decoy effect, is strong enough to make the average subject switch from being risk-averse to being risk-seeking, and that the potential decoy effect is completely absent in this environment. While other experimental studies in choice under risk have reported decoy effects, the latter finding suggests that these effects may be less robust in choice under risk \color{black} when there are more than two possible outcomes \color{black} than they are in other domains. \color{black}

\bibliographystyle{econometrica}
\bibliography{sqb&decoy}

\pagebreak

\appendix

\section*{\color{black} Appendix: Summary of the Pilot Experiment}

\begin{figure}[htb]
	\caption{\color{black} The three lotteries in the pilot experiment.}
	\label{fig:pilot_lotteries}
	\centering 
	\begin{subfigure}{0.32\textwidth}
		\includegraphics[width=\linewidth]{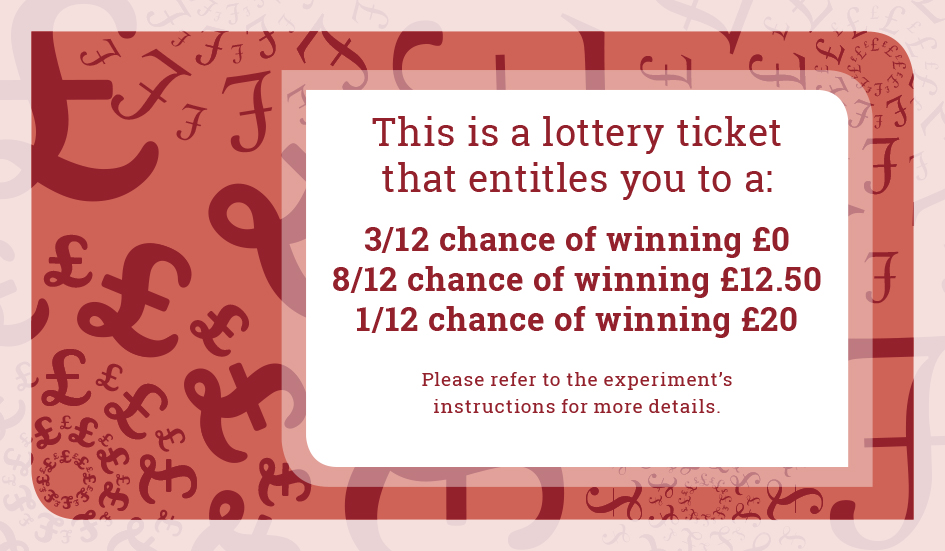}
		\caption{Lottery \textit{P}}
		\label{fig:P}
	\end{subfigure}\hfil
	\begin{subfigure}{0.32\textwidth}
		\includegraphics[width=\linewidth]{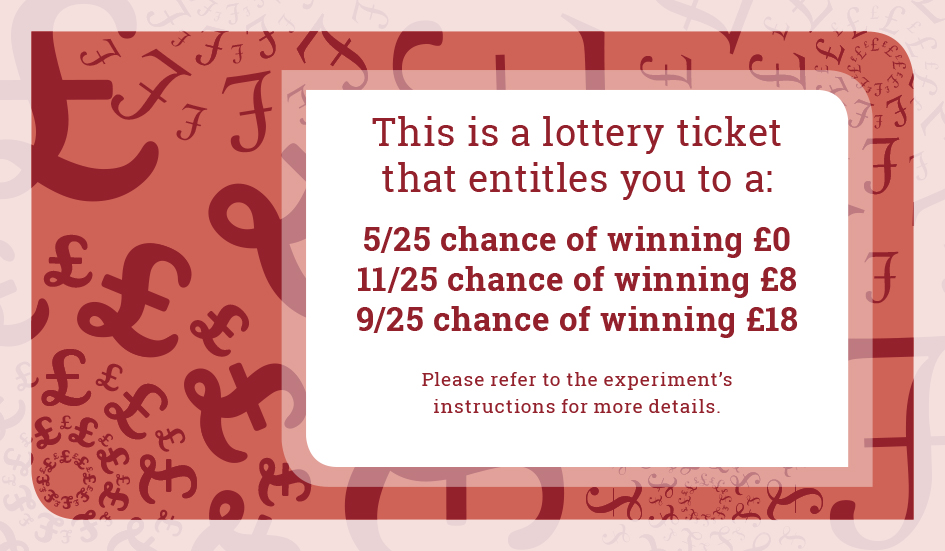}
		\caption{Lottery \textit{Q}}
		\label{fig:Q}
	\end{subfigure}\hfil 
	\begin{subfigure}{0.32\textwidth}
		\includegraphics[width=\linewidth]{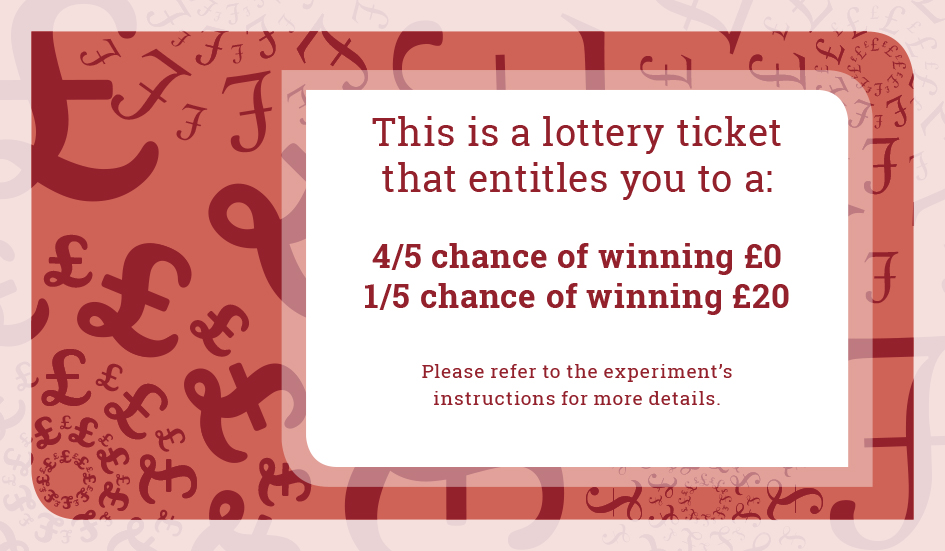}
		\caption{Lottery \textit{P-decoy}}
		\label{fig:Pdec}
	\end{subfigure}
\end{figure}

\begin{table}[!htbp]
\centering \footnotesize
\caption{\color{black} Attribute dominance within the three pairs of lotteries in the pilot experiment.}
\renewcommand{\arraystretch}{1.1}
\makebox[\textwidth][c]{
\begin{tabular}{|l|c|c|c|c|c|c|}
\hline
& \textbf{First-order}		& \textbf{Second-order}   	& \textbf{Expected}  & \textbf{Maximum} & \textbf{Probability of}   & \textbf{Probability of} \\  
& \textbf{stochastically}  	& \textbf{stochastically} 	& \textbf{value}	   & \textbf{prize}   &  \textbf{winning}   & \textbf{winning the} \\ 
& \textbf{dominant}			& \textbf{dominant}       	&					   & 				  & \textbf{more than \pounds 0}  & \textbf{lottery's}\\
\textit{Lottery pair}	& 								& 					      &					   & 				  &   &  \textbf{largest prize} \\
\hline
\textbf{\textit{P vs Q}}        & None 					& None			  & Equal		 	   & \textit{P}  	  & \textit{Q}	& \textit{Q} \\ 
\hline
\textbf{\textit{P vs P-decoy}}  & None  				& None			  & \textit{P}		   & Equal 			  & \textit{P}	& \textit{P-decoy} \\ 
\hline
\textbf{\textit{Q vs P-decoy}}  & None					& None 					  & \textit{Q} 	       & \textit{P-decoy}  	  & \textit{Q}	& \textit{P-decoy} \\
\hline
\end{tabular}
}
\label{tab:dominance_pilot}
\end{table}

\begin{table}[!htbp]
	\footnotesize
	\captionsetup{justification=centering}
	\caption{\color{black} Top panel: choice probabilities in the three treatments.\\ Bottom panel: treatment-effect tests on the choice probability of $P$.
	}
	\makebox[\textwidth][c]{
	\renewcommand{\arraystretch}{1.1}
	\begin{tabular}{|r|c|c|c|}
		\hline
		\multirow{2}{*}				& {\textbf{Treatment 1}} & \textbf{Treatment 2}	& \textbf{Treatment 3}\\ 
									& 	(binary)	 		 &  (decoy) 			& (decoy \& default)\\ 
		\hline
		\textit{P}          		&  0.34			     	 & 0.36					&  0.20 \\ 
		\hline
		\textit{Q}          		&  0.66	             	 & 0.64					&  0.80 \\
		\hline
		\textit{P-decoy}          	&  --		             &  0					&  0 \\
		\hline
		$n=$ 		          		& 35					 &	31					&  40 \\
		\hline
		$p$-value  					& 0.089					 & 0.149 				&  $<0.001$ \\
		\hline
		\hline
									& \textbf{Treatments 1 \& 2}	& \textbf{Treatments 1 \& 3} 	& \textbf{Treatments 2 \& 3}\\ 
		\hline
		$p$-value 	               	& 1	 		             		& 0.1965  						& 0.18\\
		\hline
	\end{tabular}
}
{\footnotesize Notes: (i) top-panel $p$-values from 2-tailed binomial tests with an equal-proportions null; (ii) bottom-panel $p$-values from 2-tailed Fisher's exact tests.}
	\label{tab:results_pilot}
\end{table}

\end{document}